\documentclass[11pt, fleqn]{article}

\usepackage[utf8]{inputenc}
\usepackage[T1]{fontenc}

\usepackage[a4paper,margin=2.5cm]{geometry}

\usepackage{subcaption}

\usepackage{mathtools}
\mathtoolsset{centercolon}
\numberwithin{equation}{section}
\usepackage{tensor, mathrsfs}

\usepackage[sorting=none,url=false,backend=biber,style=numeric-comp,giveninits=true]{biblatex}

\bibliography{bibliography}

\usepackage[hidelinks]{hyperref}
\usepackage{cleveref}

\usepackage{authblk}

\usepackage[onehalfspacing]{setspace}
\usepackage{lmodern}

\let\t\tensor
\let\p\partial
\def\dd{\mathrm d}
\def\o#1{{(#1)}}
\def\omegaG{\omega_g}
\def\kappaB{{\bar \kappa}}

\begin{document}
\title{The Resolution of Ambiguities in\\ Light Perturbation by Gravitational Waves}
\author[1,2]{Thomas B.~Mieling\thanks{
\textsc{orcid:} \href{https://orcid.org/0000-0002-6905-0183}{0000-0002-6905-0183},
{email:} \href{mailto:thomas.mieling@univie.ac.at}{thomas.mieling@univie.ac.at}
}}
\author[1]{Stefan Palenta\thanks{
\textsc{orcid:} \href{https://orcid.org/0000-0002-6541-9537}{0000-0002-6541-9537}}}
\affil[1]{University of Vienna, Faculty of Physics, Vienna, Austria}
\affil[2]{TURIS Research Platform, Vienna, Austria}
\date{\today}
\maketitle

\begin{abstract}
	Some previously published expressions for the perturbation of light by gravitational waves exhibit pathological behaviour in the limit of parallel propagation. We show that this is caused by similarly pathological initial or boundary data and can thus be remedied by implementing better-behaved initial conditions.
\end{abstract}

\section{Introduction}

The perturbation of light by weak gravitational waves (GW’s) has been discussed extensively in the literature \cite{Plebanski1960,Estabrook1975,Christie2007,Rakhmanov2009,Koop2014,Blaut2019}. However, some previously published results yield ill-behaved expressions in the parallel limit, where the electromagnetic and gravitational waves propagate in the same direction. Specifically, this problem arises due to the occurrence of expressions of the form
\begin{equation}
	\varPsi
		:= \frac{1}{2} \frac{\t A{^i^j} \t m{_i} \t m{_j}}{1 - \t n{^k} \t m{_k}}\,,
\end{equation}
where $m$ and $n$ are unit three-vectors (in the flat background geometry) along the light ray and the GW, respectively, and $\t A{^i^j}$ is the GW polarisation tensor (which is transverse to $n$ and traceless).
Indeed, setting
\begin{align}
	\label{eq:introduction parametrisation}
	(\t m{_i}) &=
	\begin{pmatrix}
		\cos \theta\\
		\sin \theta \cos \phi\\
		\sin \theta \sin \phi
	\end{pmatrix}\,,
	&
	(\t n{_i}) &=
	\begin{pmatrix}
		1\\
		0\\
		0
	\end{pmatrix}\,,
	&
	(\t A{^i^j}) &=
	\begin{pmatrix}
		0	&	0	&	0\\
		0	&	A_+	&	A_\times\\
		0	&	A_\times	&	- A_+
	\end{pmatrix}\,,
\end{align}
one obtains
\begin{equation}
	\lim_{\theta \to 0} \varPsi
	= \lim_{\theta \to 0} \cos^2(\theta/2)[
		A_+ \cos(2\phi)
		+ A_\times \sin(2\phi)
	]
	= A_+ \cos(2\phi)
	+ A_\times \sin(2\phi)\,.
\end{equation}
But in the limit $\theta \to 0$, the angle $\phi$ becomes meaningless so that this limit depends on the direction from which $m$ tends towards $n$. Hence, $\varPsi$ \emph{cannot} be extended continuously to $m = n$.

Such ill-behaved expressions were obtained in various perturbative calculations concerned with null geodesics \cites[Eq.~(3.15)]{Finn2009}[Sect.~III]{Angelil2015}, the electromagnetic field tensor \cites[Eq.~(2.20)]{Lobo1992}[Eq.~(3.5)]{Montanari1998}[Sect.~3]{Calura1999}, and the electromagnetic gauge potential \cites[Eq.~(2.19)]{Calura1999}[Eq.~(3.11)]{Codina1980}[Eq.~(26)]{Park2021}. (In Ref.~\cite{Calura1999} the singularity was found to be removable by a similarly singular gauge transformation.)
Here, we show that such pathological behaviour can be traced back to ill-behaved initial data of the geodesic equation, or boundary data for the eikonal equation, and can be remedied by implementing well-behaved data instead.

Specifically, we consider the perturbation of null geodesics in \Cref{s:geodesics} and perturbations of the eikonal function in \Cref{s:eikonal equation}. In both cases, we show explicitly how these ill-behaved expressions arise and give corrected solutions (constructed from well-behaved data) which are free of this pathological behaviour.

\section{Null Geodesics}
\label{s:geodesics}

Within the framework of linearised gravity where the metric $\t g{_\mu_\nu}$ is close to the Minkowski metric $(\t \eta{_\mu_\nu}) = \operatorname{diag}(-1,1,1,1)$, plane gravitational waves can be described in the standard transverse-traceless (TT) gauge as
\begin{equation}
	\t g{_\mu_\nu}
		= \t \eta{_\mu_\nu}
		+ \varepsilon \t h{_\mu_\nu}(\t \kappa{_\sigma} \t x{^\sigma})
		+ O(\varepsilon^2) \,,
\end{equation}
where $\varepsilon \ll 1$ is the GW amplitude, $\t\kappa{_\mu}$ its wave vector, and $\t h{_\mu_\nu}$ satisfies $\t h{_\mu_0} = 0$, $\t h{_\mu_\nu} \t \eta{^\nu^\rho} \t \kappa{_\rho} = 0$ as well as $\t h{_\mu_\nu} \t \eta{^\mu^\nu} = 0$.
Denoting by $\omegaG$ the GW frequency and setting
\begin{align}
	\t \kappa{_\mu} &= \omegaG(-1, + \t n{_i})\,,
	&
	\t \kappaB{_\mu} &= \omegaG(-1, - \t n{_i})\,,
\end{align}
whose indices are raised and lowered with $\t\eta{_\mu_\nu}$ ($\t n{_i}$ is the direction of GW propagation), we define the null coordinates $U$, $V$, and transverse spatial coordinates $(y^a) = (y^1, y^2)$:
\begin{align}
	\label{eq:TT null transformation}
	\omegaG U &= - \t \kappa{_\mu} \t x{^\mu}\,,
	&
	\omegaG V &= - \t \kappaB{_\mu} \t x{^\mu}\,,
	&
	(y^a)	&= \t* b{^{(a)}_i} \t x{^i}\,,
\end{align}
where $(n, \t b{^{(1)}}, \t b{^{(2)}})$ is any (fixed) orthonormal basis of Euclidean space (the time components of $n$ and $\t b{^{(a)}}$ are set to zero).
The metric then takes the form
\begin{equation}
	\label{eq:metric Einstein Rosen approx}
	g = - \dd U\, \dd V
		+ (\t \delta{_a_b} + \varepsilon \t h{_a_b}(- \omegaG U))\, \dd\t y{^a}\, \dd\t y{^b}
		+ O(\varepsilon^2) \,,
\end{equation}
which can be regarded as an approximate version of a Rosen metric
\begin{equation}
	\label{eq:metric Einstein Rosen}
	g = - \dd U\, \dd V
		+ \t g{_a_b}(U)\, \dd\t y{^a}\, \dd\t y{^b}\,,
\end{equation}
which also allows for solutions to the full nonlinear Einstein vacuum equations \cites[§~109]{Landau1994}[Sect.~24.5]{Stephani2003}{Cropp2010}.
Since the computation of light rays in the metric \eqref{eq:metric Einstein Rosen} is no more difficult than in \eqref{eq:metric Einstein Rosen approx}, we consider the exact case first and specialise to the perturbative setting afterwards.

\subsection{Non-Perturbative Solution}
\label{s:geodesics non-perturbative}

Using the obvious Killing vector fields of \eqref{eq:metric Einstein Rosen}, one obtains the constant momenta
\begin{align}
	\t p{_V} &= - \tfrac{1}{2} \dot U\,,
	&
	\t p{_a} &= \t g{_a_b}(U) \t{\dot y}{^a}\,.
\end{align}
Integrating the equations $\t p{_V} = \text{const.}$ and $\t p{_a} = \text{const.}$ determines $U$ and $\t y{^a}$ as a function of 
the affine parameter $s$. Finally, $V(s)$ is obtained from the equation $4 \t p{_U} \t p{_V} = \t g{^a^b} \t p{_a} \t p{_b}$, leading to the result
\begin{align}
	\label{eq:rosen U}
	U(s)
		&= U(0) - 2 \t p{_V} s\,,\\
	\label{eq:rosen V}
	V(s) 
		&= V(0) + \frac{1}{2 \t p{_V}} \int_0^s \t g{^a^b}(U(s')) \dd s'\, \t p{_a} \t p{_b} \, \,,\\
	\label{eq:rosen a}
	\t y{^a}(s)
		&= \t y{^a}(0) + \int_0^s \t g{^a^b}(U(s')) \,\dd s'\, \t p{_b}\,,
\end{align}
cf.\ Ref.~\cite{Poplawski2006}. (We assume $\t p{_V} \neq 0$ and consider the limiting case afterwards.)
Defining
\begin{equation}
	\t G{^\mu^\nu}(s)
		= \int_0^s \t g{^\mu^\nu}(U(s)) \, \dd s'\,,
\end{equation}
the solution can be written concisely as
\begin{equation}
	\label{eq:rosen displacement}
	\t x{^\mu}(s)
		= \t x{^\mu}(0)
		+ \left(
			\t G{^\mu^\nu}(s) \t \kappa{^\sigma} - \tfrac{1}{2} \t\kappa{^\mu} \t G{^\nu^\sigma}(s)
		\right) \frac{\t p{_\nu}(0) \t p{_\sigma}(0)}{\t \kappa{^\lambda} \t p{_\lambda}(0)}\,,
\end{equation}
where $\t p{_\mu}(0)$ is the initial momentum.
This equation is readily verified by noting that \cref{eq:rosen U} is obtained by contracting with $\t \kappa{_\mu}$, \cref{eq:rosen V} by contraction with $\t \kappaB{_\mu}$ and \cref{eq:rosen a} by contraction with $\t b{^{(a)}_\mu}$.
Note that this equation is covariant under linear coordinate transformations.

\subsection{Perturbative Solution}
\label{s:geodesics perturbative}

Consider now the perturbative setting in TT coordinates:
\begin{equation}
	\label{eq:TT metric inverse}
	\t g{^\mu^\nu} = \t \eta{^\mu^\nu} - \varepsilon \t h{^\mu^\nu}(U) + O(\varepsilon^2)\,.
\end{equation}
Since the coordinate transformation given in \eqref{eq:TT null transformation} is linear, \cref{eq:rosen displacement} also applies to the TT coordinates.
Inserting \eqref{eq:TT metric inverse} into \eqref{eq:rosen displacement} and expanding the initial momentum as $\t p{_\mu}(0) = \t*p{^{\o0}_\mu} + \varepsilon \t*p{^{\o1}_\mu} + O(\varepsilon^2)$, one obtains the perturbed light rays
\begin{align}
	\label{eq:ray expansion}
	\t x{^\mu}(s)
		&= \t x{^\mu}(0)
		+ s\, \t \eta{^\mu^\nu} \t*p{^{\o0}_\nu}
		+ \varepsilon \t x{^{\o1}^\mu}(s)
		+ O(\varepsilon^2)\,,
	\\
	\shortintertext{where}
	\label{eq:ray perturbation}
	\t x{^{\o1}^\mu}(s)
		&= s\, \t \eta{^\mu^\nu} \t*p{^{\o1}_\nu}
		- \bigg(
			\tfrac{1}{2} s \t \kappa{^\mu} \t h{^\nu^\sigma}(u_0)
			+ \t H{^\mu^\nu}(s) \t \kappa{^\sigma}
			- \tfrac{1}{2} \t \kappa{^\mu} \t H{^\nu^\sigma}(s)
		\bigg) \frac{\t*p{^{\o0}_\nu} \t*p{^{\o0}_\sigma}}{ \t\kappa{^\lambda} \t*p{^{\o0}_\lambda}}\,,
	\\
	\label{eq:integrated waveform}
	\t H{^\mu^\nu}(s)
		&= \int_0^s \t h{^\mu^\nu}(U^\o0(s)) \,\dd s'\,,
\end{align}
where $u_0 = - \t \kappa{_\rho} \t x{^\rho}(0)/\omega$. Here, we have used $(\t \eta{^\mu^\nu} - \varepsilon \t h{^\mu^\nu}(u_0)) \t p{_\mu}(0) \t p{_\nu}(0) = O(\varepsilon^2)$.

While the last two terms in \cref{eq:ray perturbation} are uniformly bounded in $s$ in typical applications (for monochromatic waves, $\t H{^\mu^\nu}$ is oscillatory with zero average), the first two terms grow linearly with the affine parameter $s$, a behaviour which was interpreted as an indication of the breakdown of the perturbative expansion in Ref.~\cite{Finn2009}. These terms can be eliminated by an appropriate choice of initial momentum:
\begin{equation}
	\label{eq:momentum tentative}
	\t*p{^*_\mu}
		= \t*p{^{\o0}_\mu}
		+ \tfrac{1}{2} \varepsilon \t \kappa{_\mu} \t h{^\nu^\sigma}(u_0) \frac{\t*p{^{\o0}_\nu} \t*p{^{\o0}_\sigma}}{ \t\kappa{^\lambda} \t*p{^{\o0}_\lambda}}
		+ O(\varepsilon^2)\,,
\end{equation}
which leads to the following \emph{tentative} first-order correction of the trajectory:
\begin{equation}
	\label{eq:curve tentative}
	\t*x{_*^{\o1}^\mu}
		=
		\bigg(
			\tfrac{1}{2} \t \kappa{^\mu} \t H{^\nu^\sigma}(s)
			- \t H{^\mu^\nu}(s) \t \kappa{^\sigma}
		\bigg) \frac{\t*p{^{\o0}_\nu} \t*p{^{\o0}_\sigma}}{ \t\kappa{^\lambda} \t*p{^{\o0}_\lambda}}\,,
\end{equation}
which was also obtained in Ref.~\cite{Finn2009} using renormalisation techniques. However, this expression has a pathological parallel limit (recall that we have excluded the case $\t p{_V} = 0$).
To see this, consider a polarised monochromatic GW with wave vector $\t \kappa{_\mu} = \omegaG(-1, \t n{_i})$ and waveform ${\t h{^a^b}= \t A{^a^b} \cos(\omegaG U)}$. Choosing the initial momentum $\t*p{^{\o0}_\mu} = (-1, \t m{_i})$, and using the parametrisation \eqref{eq:introduction parametrisation}, one finds
\begin{equation}
	\t H{^a^b}(s)
		=  \t A{^a^b} \frac{\sin(\omegaG s (1 - \cos \theta))}{\omegaG (1 - \cos \theta)}\,,
\end{equation}
which tends to $s \t A{^a^b}$ as $\theta \to 0$.
In this limit one obtains
\begin{equation}
	\lim_{\theta \to 0} \t*x{_*^{\o1}^\mu}
		= - s\, \t\kappa{^\mu} \omegaG^{-1} [
			A_+ \cos(2 \phi) + A_\times \sin(2\phi)
		]\,,
\end{equation}
which depends on the geometrically undefined angle $\phi$. This is the pathological behaviour described in the introduction.

In fact, this can be traced back entirely to the similarly pathological behaviour of the initial momentum $p^*$:  inserting the above parametrisation into \cref{eq:momentum tentative} one finds
\begin{equation}
	\lim_{\theta \to 0} \t*p{^*_\mu}
		= \t \kappa{_\mu} \omegaG^{-1}
		\left( 1 - \varepsilon [h_+(u_0) \cos(2 \phi) + h_\times(u_0) \sin(2\phi)] \right)
		+ O(\varepsilon^2)\,,
\end{equation}
which also depends on the geometrically undefined angle $\phi$.
It is thus clear that the problematic behaviour of \cref{eq:curve tentative} can be cured only by choosing a different initial momentum.

The fact that the parallel limit yields the correct momentum only up to an ill-behaved multiplicative factor suggests that the problem can be remedied by changing the parametrisation of the curve. This is indeed the case: adding a multiple of $\t*p{^{\o0}_\nu}$ to $\t*p{^*_\nu}$ such that the overall momentum satisfies $\t p{_0}(0) = -1$, say, one is led to the \emph{improved} initial momentum
\begin{equation}
	\label{eq:momentum improved}
	\t*p{^\dagger_\mu}
		= \t*p{^{\o0}_\mu}
		+ \tfrac{1}{2} \varepsilon (\t \kappa{_\mu} - \omegaG \t*p{^{\o0}_\mu}) \t h{^\nu^\sigma}(0) \frac{\t*p{^{\o0}_\nu} \t*p{^{\o0}_\sigma}}{ \t\kappa{^\lambda} \t*p{^{\o0}_\lambda}}
		+ O(\varepsilon^2)\,,
\end{equation}
which \emph{is} well-behaved in the collinear limit: $\lim_{\theta \to 0} \t*p{^\dagger_\mu} = \t*p{^{\o0}_\mu}$.
With this choice, the first-order correction of the light ray now takes the form
\begin{equation}
	\label{eq:curve improved}
	\t x{^{\o1}^\mu}
		= \bigg(
			\tfrac{1}{2} \t \kappa{^\mu} \t H{^\nu^\sigma}(s)
			- \t H{^\mu^\nu}(s) \t \kappa{^\sigma}
			- \tfrac{1}{2} \omegaG s\, \t p{^{\o0}^\mu} \t h{^\nu^\sigma}(0)
		\bigg) \frac{\t*p{^{\o0}_\nu} \t*p{^{\o0}_\sigma}}{ \t\kappa{^\lambda} \t*p{^{\o0}_\lambda}}\,.
\end{equation}
One readily verifies that this expression vanishes for $\theta \to 0$: as one would expect, light rays are not perturbed by co-propagating gravitational waves. Because the linearly growing term here is proportional to the unperturbed tangent vector, it \emph{does not} describe a physical separation from the unperturbed ray: the physical first-order perturbation of the curve always remains bounded whenever $\t H{^\mu^\nu}$ does.

While the construction of $\t*p{^\dagger_\mu}$  given here might appear ad-hoc, it resolves the issues explained in the introduction. It is also characterised geometrically by the following three conditions: (i) $p^\dagger$ is lightlike; (ii) $\t*p{^\dagger_0} = -1$ so that the initial energy is normalised; (iii) there is a constant $\lambda$ such that for all three Killing vectors $K = \partial_v, \partial_y, \partial_z$ one has $\t*p{^\dagger_\mu} \t K{^\mu} = \lambda \t*p{^{\o0}_\mu} \t K{^\mu}$.


\begin{figure}[ht]
	\centering
	\begin{subfigure}[c]{0.44\textwidth}
		\includegraphics[width=\columnwidth]{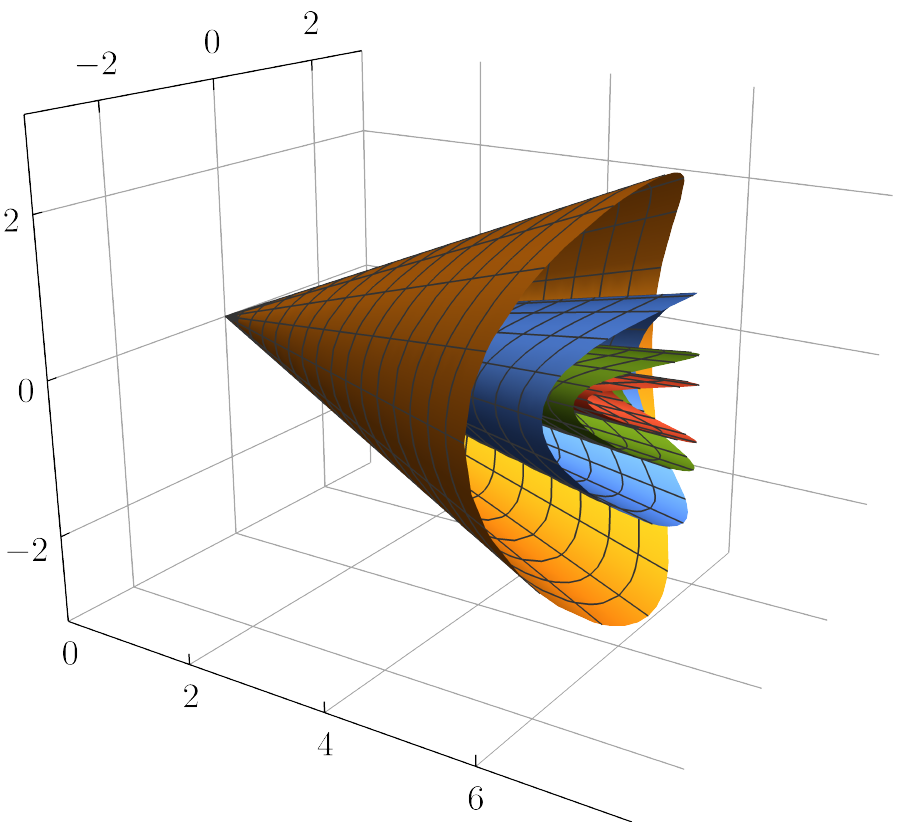}
		\subcaption{Pathological light cones computed from the curves \eqref{eq:curve tentative}.}
		\label{fig:cone singular}
	\end{subfigure}
	\hspace{0.1\textwidth}
	\begin{subfigure}[c]{0.44\textwidth}
		\includegraphics[width=\columnwidth]{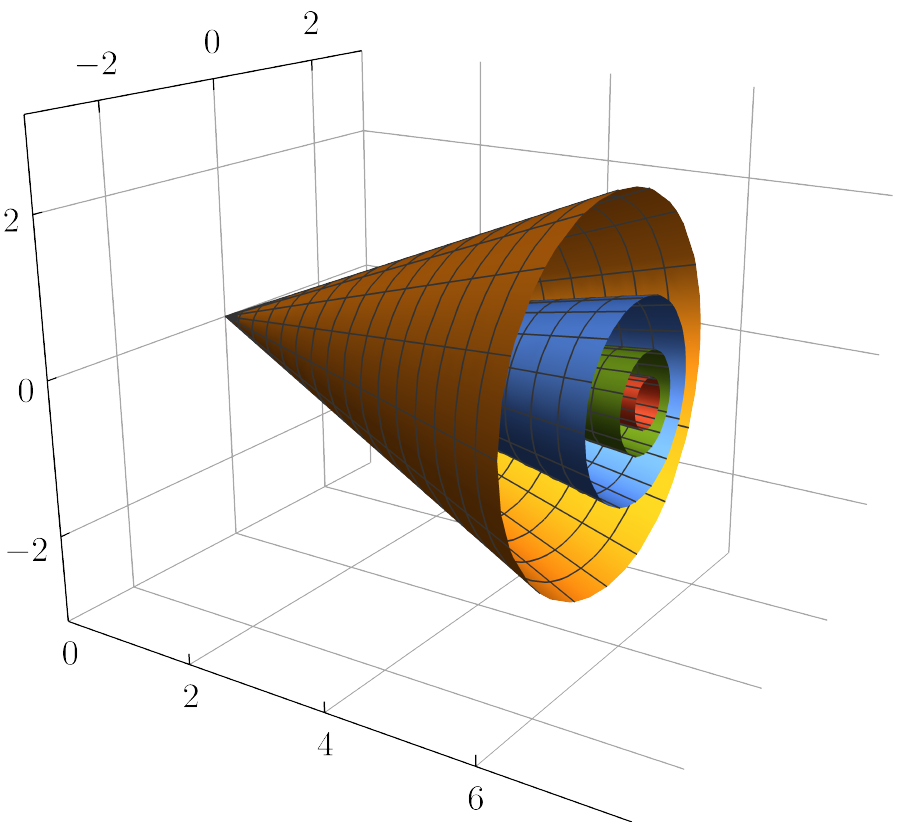}
		\subcaption{Regular light cones computed from the curves \eqref{eq:curve improved}.}
		\label{fig:cone regular}
	\end{subfigure}
	\caption{Spatial light cones of decreasing opening angle.
	The left panel (a) demonstrates the direction-dependence of the limit $\theta \to 0$ of \cref{eq:curve tentative}: coming from above, one covers a larger coordinate distance than coming from the side. Contrarily, the cones constructed from \cref{eq:curve improved} are free of this ambiguity (b).
	For visualisation, we have set $\omegaG = 1$, $\varepsilon = 0.1$ and have drawn the light rays up to an affine parameter of $s = 2\pi$.
	}
	\label{fig:cones}
\end{figure}

\Cref{fig:cones} shows \emph{spatial} light cones of decreasing opening angle obtained by varying $\phi$ over the range $(0, 2 \pi)$ and varying the affine parameter $s$ up to one GW wavelength.
\Cref{fig:cone singular} shows the cones obtained from \cref{eq:curve tentative} which does not extend continuously to the parallel case: when the opening angle is shrunk to zero, the covered coordinate distance depends on the direction from which the limit is taken.
Conversely, \cref{eq:curve improved} is free of this issue, as shown in \Cref{fig:cone regular}.

\begin{figure}[ht]
	\centering
	\begin{subfigure}[c]{0.44\textwidth}
		\includegraphics[width=\columnwidth]{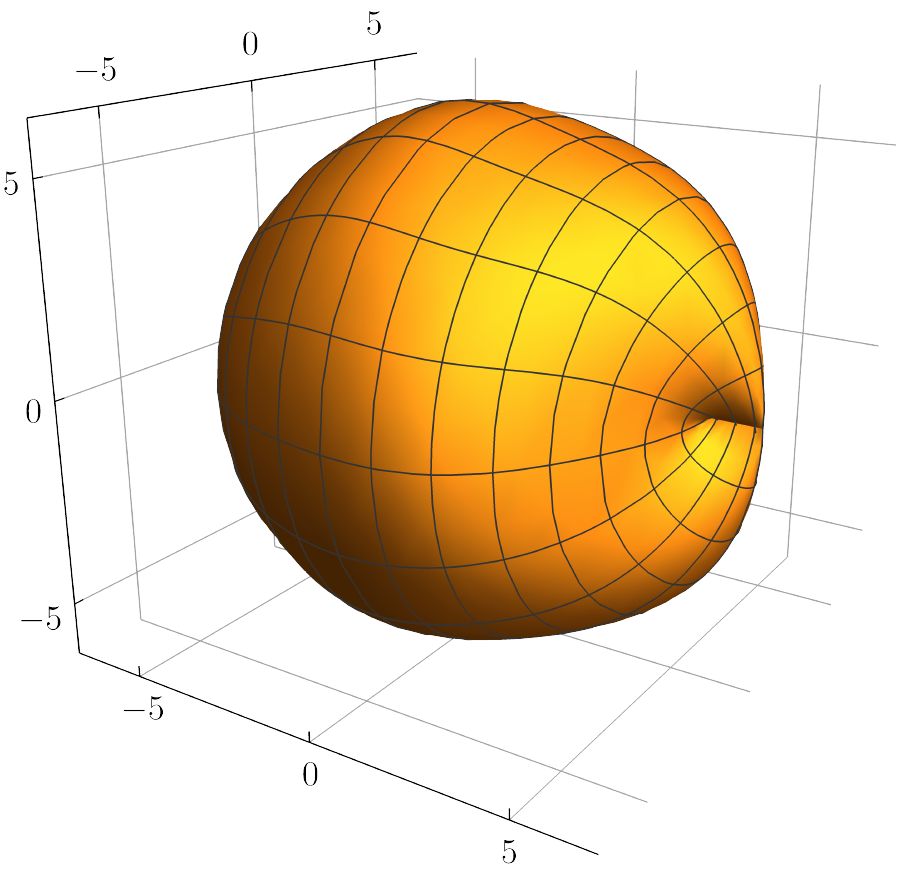}
		\subcaption{Pathological light sphere computed from the curves \eqref{eq:curve tentative}.}
		\label{fig:sphere singular}
	\end{subfigure}
	\hspace{0.1\textwidth}
	\begin{subfigure}[c]{0.44\textwidth}
		\includegraphics[width=\columnwidth]{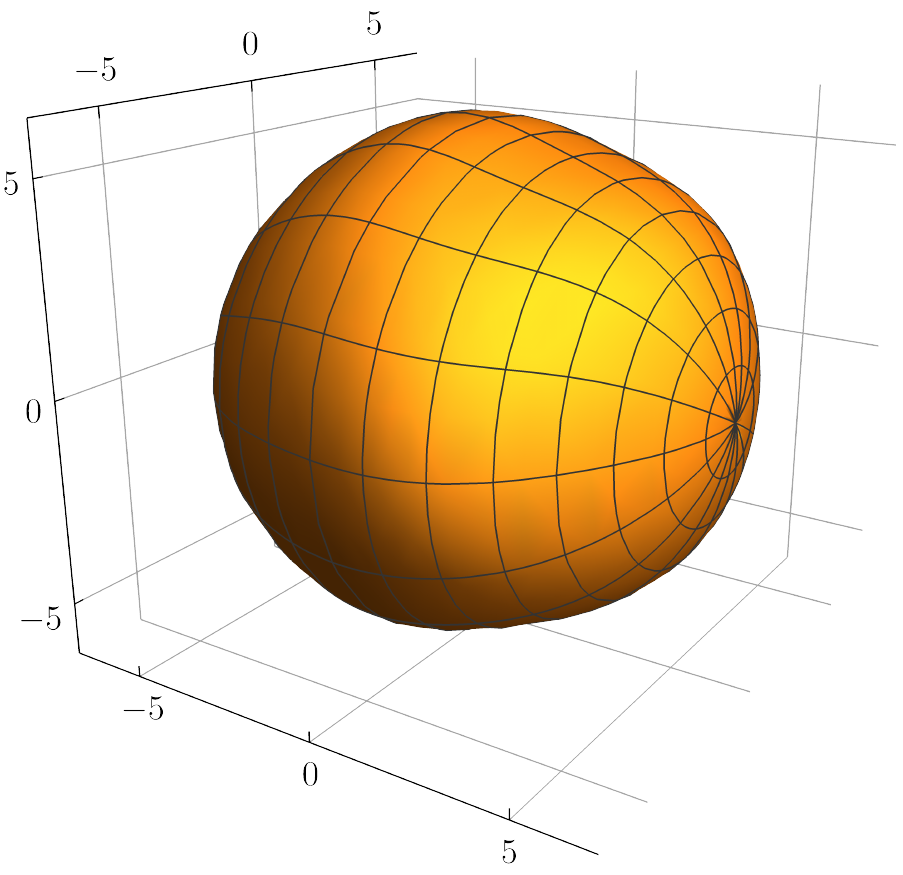}
		\subcaption{Regular light sphere computed from the curves \eqref{eq:curve improved}.}
		\label{fig:sphere regular}
	\end{subfigure}
	\caption{“Light Spheres” obtained by following light rays from a point along all possible directions for an affine parameter of $s = 2\pi$ (with $\omegaG = 1$ and $\varepsilon = 0.1$ for visualisation). The discontinuity of \cref{eq:curve tentative} produces a “fold” in the direction of GW propagation (a), while the sphere constructed from \cref{eq:curve improved} is continuous there (b).}
	\label{fig:sphere}
\end{figure}

This discontinuity can be visualised also by constructing “spheres” by varying the emission angles $\theta, \phi$ while keeping the affine parameter, $s$, fixed. The resulting (spatial) wave fonts are shown in \Cref{fig:sphere}. The discontinuity of \cref{eq:curve tentative} results in a “notch”, see \Cref{fig:sphere singular}, while \cref{eq:curve improved} gives rise to the regular surface shown in \Cref{fig:sphere regular}.

As the two curves corresponding to the initial momenta $p^*$ and $p^\dagger$ are equivalent in the sense that one can be obtained from the other by re-parametrisation, their predictions for the light trajectories coincide. However, care must be taken when the affine parameter is interpreted as an optical phase (as is the case in geometrical optics). Indeed, similar pathologies arise in the eikonal equation, as discussed in the next section.

\section{Eikonal Equation}
\label{s:eikonal equation}

One of the fundamental quantities of geometrical optics is the eikonal function $\psi$, satisfying the eikonal equation
\begin{equation}
	\label{eq:eikonal equation}
	\t g{^\mu^\nu} (\t\nabla{_\mu} \psi) (\t\nabla{_\nu} \psi) = 0\,.
\end{equation}
In this description, light rays are modelled by integral curves of $\t\nabla{^\mu} \psi$, which are null geodesics along which the eikonal is constant.
To describe the perturbation of plane light waves, consider perturbative solutions of the form
\begin{equation}
	\psi = \t*k{^{\o0}_\mu} \t x{^\mu} + \varepsilon \psi^\o1 + O(\varepsilon^2)\,,
\end{equation}
where $\t*k{^{\o0}_\mu}$ is a constant null vector in the flat background geometry.
Inserting this into \cref{eq:eikonal equation} with the metric given in \cref{eq:TT metric inverse} and neglecting terms of order $\varepsilon^2$, one obtains
\begin{equation}
	\label{eq:eikonal transport}
	\t*k{^{\o0}^\mu} \t\p{_\mu} \psi^\o1 = \tfrac{1}{2} \t h{^\mu^\nu}(\t \kappa{_\sigma} \t x{^\sigma}) \t*k{^{\o0}_\mu} \t*k{^{\o0}_\mu}\,,
\end{equation}
where $\t*k{^{\o0}^\mu} = \t\eta{^\mu^\nu}\t*k{^{\o0}_\nu}$.
For a unique solution one must prescribe boundary values on a suitable hypersurface $\Sigma$, for which we introduce some notation. Given any point $\t x{^\mu}$, denote by $\t* x{_{\text{ret.}}^\mu}$ the point obtained by following the unperturbed light rays back in time (along $- \t k{^{\o0}^\mu}$) until one reaches $\Sigma$.
This can be written as
\begin{equation}
	\t*x{_{\text{ret.}}^\mu} = \t x{^\mu} - s(x) \t k{^{\o0}^\mu}\,,
\end{equation}
where the “parametric distance” $s$ satisfies
\begin{align}
	\t k{^{\o0}^\mu} \t\p{_\mu} s &= 1\,,
	&
	s |_\Sigma = 0\,.
\end{align}
The solution to \cref{eq:eikonal transport} with prescribed boundary data $\psi^\o1 |_\Sigma = \delta \psi$ can then be written as
\begin{equation}
	\psi^\o1
		= \frac{\t k{^{\o0}_\mu} \t k{^{\o0}_\mu}}{2 \t \kappa{^\lambda} \t k{^{\o0}_\lambda}} \int_{\t \kappa{_\sigma} \t*x{_{\text{ret.}}^\sigma}}^{\t \kappa{_\sigma} \t x{^\sigma}} \t h{^\mu^\nu}(u)\, \dd u
		+ \delta \psi(x_{\text{ret.}})\,.
\end{equation}
This is the general expression for the eikonal perturbation for arbitrary boundary values and GW waveforms.
To get some insight into this equation, consider a monochromatic polarised GW with $\t h{^\mu^\nu} = \t A{^\mu^\nu} \cos(U)$. In this case, one obtains
\begin{equation}
	\psi^\o1
		= \frac{\t A{^\mu^\nu}\t k{^{\o0}_\mu} \t k{^{\o0}_\mu}}{2 \t \kappa{^\lambda} \t k{^{\o0}_\lambda}} 
		[
			\sin(\t \kappa{_\sigma} \t x{^\sigma})
			- \sin(\t \kappa{_\sigma} \t x{^\sigma} + s(x) \t \kappa{_\sigma} \t k{^{\o0}^\sigma})
		]
		+ \delta \psi(x - s k^\o0)\,.
\end{equation}
Further, setting $\t*k{^{\o0}_\mu} = \omega (-1, \t m{_i})$ for some constant vector $\t m{_i}$ (normalised in the flat background geometry), and choosing $\Sigma = \{ \t m{_i} \t x{^i} = 0\}$ (on which the phase is spatially constant in the unperturbed case), one has $s = \t m{_i} \t x{^i} / \omega$, and the eikonal perturbation takes the form
\begin{equation}
	\label{eq:eikonal perturbation plane GW}
	\psi^\o1
		= - \frac{\omega \t A{^i^j}\t m{_i} \t m{_j}}{2 \omegaG (1 - \t n{^k} \t m{_k})} 
		[
			\sin(\t \kappa{_\sigma} \t x{^\sigma})
			- \sin(\t \kappa{_\sigma} \t x{^\sigma} -  \omegaG \t m{_l} \t x{^l} (1 - \t n{^k} \t m{_k}))
		]
		+ \delta \psi(x - s k^\o0)\,.
\end{equation}
The solutions given in Refs.~\cites[Eq.~(40)]{Angelil2015}[Eq.~(26)]{Park2021} and \cite[Eq.~(2.19)]{Lobo1992} (using a further unidimensional approximation) correspond to the following choice of boundary values:
\begin{equation}
	\label{eq:eikonal bdry pathological}
	\delta \psi_*
		= - \frac{\omega \t A{^i^j}\t m{_i} \t m{_j}}{2 \omegaG (1 - \t n{^k} \t m{_k})} \sin(\t \kappa{_\sigma} \t x{^\sigma}) \bigg|_\Sigma\,,
\end{equation}
so that the last two terms in \cref{eq:eikonal perturbation plane GW} cancel. This leads to the expression
\begin{align}
	\label{eq:eikonal pathological}
	\psi^\o1_*
		= - \frac{\omega \t A{^i^j}\t m{_i} \t m{_j}}{2 \omegaG (1 - \t n{^k} \t m{_k})}
		\sin(\t \kappa{_\sigma} \t x{^\sigma})\,,
\end{align}
which has precisely the pathological behaviour described in the introduction.
As for single light rays, this is caused by the pathological behaviour of the boundary data \eqref{eq:eikonal bdry pathological}: it has the same form as \eqref{eq:eikonal pathological} and thus suffers from the same problems.

This issue is resolved by choosing well-behaved boundary data instead. As explained in detail in Refs.~\cite{Mieling_2021b,Mieling:2021c}, normal emission of light from $\Sigma$ with constant frequency $\omega$ corresponds to $\delta \psi \equiv 0$, which yields instead
\begin{equation}
	\psi^\o1
		= - \frac{\omega \t A{^i^j}\t m{_i} \t m{_j}}{2 \omegaG (1 - \t n{^k} \t m{_k})} 
		[
			\sin(\t \kappa{_\sigma} \t x{^\sigma})
			- \sin(\t \kappa{_\sigma} \t x{^\sigma} -  \omegaG \t m{_l} \t x{^l} (1 - \t n{^k} \t m{_k}))
		]\,.
\end{equation}
Inserting the parametrisation \eqref{eq:introduction parametrisation}, it is readily verified that the parallel limit of this expression is well-behaved: $\lim_{\theta \to 0}\psi^\o1 = 0$, which is in agreement with the expectation light rays are not perturbed by co-propagating gravitational waves.

\subsection{Inadequacy of Separation of Variables}

Given the simple solution of the geodesic equation in \Cref{s:geodesics non-perturbative}, one might hope for a similarly simple solution of the eikonal equation
\begin{equation}
	-4 (\t\p{_u} \psi) (\t\p{_v} \psi) + \t g{^a^b}(u) (\t\p{_a} \psi) (\t\p{_b} \psi) = 0\,,
\end{equation}
without approximations.
Indeed, the standard method of separation of variables (as commonly used in the Hamilton-Jacobi formulation of mechanics) directly leads to the complete integral
\begin{equation}
	\psi
		= \t p{_V} V + \t p{_a} \t y{^a} + \frac{\t p{_a} \t p{_b}}{4 \t p{_V}} \int \t g{^a^b}(u) \,\dd u\,,
\end{equation}
with constants $\t p{_V}$ and $\t p{_a}$, as also found in Ref.~\cite[Eq.~(7)]{Poplawski2006}.
Expanding the metric tensor $\t g{^a^b}$ to first order in $\varepsilon$, one obtains
\begin{align}
	S
		&= S^\o0 + \varepsilon S^\o1 + O(\varepsilon^2)\,,
		\\
	\shortintertext{where}
	S^\o0
		&= \t p{_V} V + \t p{_a} \t y{^a} + \tfrac{1}{4} u \t{\delta}{^a^b}(u) \t p{_a} \t p{_b} / \t p{_V}\,,
		\\
	S^\o1
		&= - \frac{\t p{_a} \t p{_b}}{4 \t p{_V}} \int \t h{^a^b}(u) \,\dd u\,.
\end{align}
Evidently, $S^\o0$ satisfies the Eikonal equation in flat space and has a well-behaved parallel limit.
However, the parallel limit of $S^\o1$ is ill-behaved, as it is of the same form as discussed in the previous section.
Thus, the method of separation of variables is inadequate to compute the eikonal for arbitrary GW incidence angles.

\section{Conclusion}

We have noticed that some previously published expressions for the perturbations of light rays and phase functions, due to gravitational waves, exhibit pathological behaviour in the limit where light is emitted in parallel to the gravitational wave. 
Contrary to the interpretation that this issue signals a break-down of perturbative expansions \cite{Finn2009} or even a limit in the range of validity of geometrical optics \cite{Park2021}, our analysis shows that this is due the use of ad-hoc particular solutions of the equations without checking their physical plausibility.

While we have restricted the discussion to single light rays and the eikonal equation for simplicity of exposition, the argument carries over to more accurate models based on wave equations (specifically those implied by Maxwell’s equations). As shown in Ref.~\cite{Mieling:2021c}, a careful choice of boundary data for Maxwell’s equations is necessary to obtain similarly well-behaved solutions to the full electromagnetic field emitted from a radiating surface.

\section*{Acknowledgements}

We thank Piotr Chruściel for helpful discussions.
Research supported in part by the Austrian Science Fund (FWF), Project P34274 and Grant TAI 483-N, as well as the Vienna University Research Platform TURIS. TBM is a recipient of a DOC Fellowship of the Austrian Academy of Sciences at the Faculty of Physics of the University of Vienna and acknowledges partial support from the Vienna Doctoral School in Physics (VDSP).

\singlespacing\footnotesize
\printbibliography
\end{document}